\documentclass[11pt]{article}
\usepackage[margin=1in]{geometry}
\usepackage{amsmath,amssymb}
\usepackage[super,sort&compress]{natbib}
\usepackage{booktabs}
\usepackage{array}
\usepackage{graphicx}
\usepackage{float}
\usepackage{needspace}
\usepackage{xcolor}
\newcommand{\rev}[1]{{#1}}
\newcommand{\revisioncolor}{}
\usepackage[hidelinks]{hyperref}
\usepackage{microtype}
\usepackage{indentfirst}
\usepackage{fancyhdr}
\newcommand{\PF}{\mathrm{PF}}

\title{\bfseries Bayesian Optimization for Dose Finding with Two Agents: Participant Allocation and Final Selection}
\author{Xinzhu Wang \quad Tanzy Love\\[2pt]
  \small Department of Biostatistics and Computational Biology, University of Rochester\\
  \small \texttt{xinzhu\_wang@urmc.rochester.edu}; \texttt{tanzy\_love@urmc.rochester.edu}}
\date{}

\begin{document}
\maketitle

\begin{abstract}
\rev{\noindent In two-agent dose-finding trials, the next cohort should help identify a combination for final selection. We studied a constrained knowledge-gradient (cKG) rule with one-cohort lookahead that updates independent Gaussian-process models of efficacy and continuous toxicity, reapplies a probability criterion for mean toxicity, and evaluates the resulting selection. We derived a deterministic calculation over a fixed set of dose combinations, holding fitted model parameters fixed during each hypothetical update. We compared cKG with constrained expected improvement (cEI) and two toxicity-only rules, targeted mean squared error (tMSE) and entropy, in four synthetic scenarios. In the primary obstructive sleep apnea (OSA)-derived scenario, averaged equally over strata and five probability cutoffs, cKG assigned fewer participants to combinations above the true mean-toxicity limit than tMSE (17.92\% versus 27.08\%), but selected such combinations more often at trial completion (18.80\% versus 11.85\%). Compared with cEI, cKG had higher mean simulated reduction in the 4\%-desaturation apnea--hypopnea index (AHI4) at final selection (7.46 versus 6.72 events/hour), more above-limit final selections (18.80\% versus 10.50\%), and more above-limit assignments (17.92\% versus 15.10\%). Across scenarios, its efficacy advantage over cEI was smaller under stricter toxicity criteria. Continuous outcomes, uncalibrated toxicity limits, and a rule that still selects a combination when none meets the criterion limit clinical interpretation. Allocation and final-selection toxicity should be reported separately, alongside efficacy.}
\end{abstract}


\section{Introduction}

Dose finding with two agents requires a choice among combinations that may differ in both efficacy and toxicity. An optimal biological dose (OBD) is often defined through a benefit--risk criterion rather than toxicity alone \citep{thall2004,kessels2026}. \rev{We maximize mean efficacy subject to an upper limit on mean toxicity.} Two-agent dose combinations do not form a single ordered sequence: increasing one agent while decreasing the other \rev{need not reduce toxicity}, and efficacy may plateau or decline. With a small trial and many dose pairs, most combinations receive little or no direct observation \citep{riviere2015,willard2025}. We ask which eligible combination should receive the next cohort to inform the dose selected at trial completion.

\rev{Bayesian optimization (BO) combines a response model with a rule for choosing the next dose combination. Gaussian-process (GP) models use an assumed covariance structure to predict responses at other combinations from the observed outcomes. The allocation rule, or acquisition function, scores eligible combinations according to predicted efficacy, uncertainty about toxicity, or the expected effect of new observations on the final choice.}

Existing dose-optimization designs address a broader set of decisions. BOIN12 and Comb-BOIN12 use utility-based, model-assisted rules; DROID identifies a therapeutic dose range before randomized comparison; and BARD combines backfill with adaptive randomization \citep{lin2020,lu2025combboin,guo2023,zhao2025}. \rev{Penalized D-optimality has also been used} to learn a combination-toxicity surface while respecting admissibility and escalation rules \citep{ajimi2026}. \rev{We hold the response models, assignment restrictions, and final-selection rule fixed to compare the objectives used to allocate the next cohort.}

BO and information-directed allocation already have dose-finding applications \citep{domenicano2019,takahashi2021,willard2025}. \rev{The combination-dose designs of Willard et al. use constrained expected improvement (cEI), weighting potential efficacy improvement by the probability of meeting a toxicity limit} \citep{willardobd,willardobd2026}. \rev{Their updated design uses fully Bayesian GP fitting, dose escalation, and stopping based on uncertainty about the optimal dose} \citep{willardobd2026}. \rev{We adapt the knowledge-gradient (KG) approach} \citep{frazier2009,scott2011,chen2021,ungredda2024} to ask how one additional cohort would change the best posterior mean efficacy among combinations meeting a toxicity criterion. Both response models are updated before that criterion is reapplied. \rev{The resulting constrained KG (cKG) score is positive when the expected value of the final selection increases, and can be negative when toxicity learning removes promising combinations from consideration. We derive a deterministic evaluator for a finite grid with independent Gaussian response models for efficacy and toxicity.} \rev{We compare cKG with cEI on the same OSA-derived surface used by Willard et al. and on three additional surfaces.} Two toxicity-only acquisitions, targeted mean squared error (tMSE) and an entropy-reduction rule, provide further comparators.

\rev{Our cKG formulation differs from existing constrained KG methods in its final-selection rule. Ungredda and Branke maximize posterior efficacy weighted by feasibility probability, corresponding to a specified penalty for an infeasible terminal choice \citep{ungredda2024}. Chen et al. combine expected improvement in the best constraint-mean-feasible prediction with query-feasibility weighting \citep{chen2021}. Here, a probability cutoff defines the selection set; if it is empty, a standardized-toxicity fallback determines the choice. The cKG score is the expected change in the value of this rule.}

\rev{Obstructive sleep apnea (OSA) illustrates the trade-off between efficacy and tolerability.} \rev{In MARIPOSA, placebo-adjusted changes in the 4\%-desaturation apnea--hypopnea index (AHI4) were similar for the 2.5/75- and 5/75-mg aroxybutynin/atomoxetine groups: $-7.16$ and $-7.20$ events/hour.} Dry mouth was reported in 24\% and 59\%, respectively. The later 26-week SynAIRgy trial evaluated the 2.5/75-mg pair \citep{schweitzer2023,strollo2026}. \rev{These fixed-arm trials motivate the dose-choice problem, although their data do not identify a two-dimensional dose-response surface.} Our primary scenario uses \rev{the synthetic OSA model of Willard et al.} \citep{willardobd}. \rev{We also use an author-specified MARIPOSA-motivated surface, informed qualitatively by the published results, and generic logistic and Gaussian-bump surfaces.} The logistic scenario uses a monotone dose--toxicity geometry familiar from early-phase dose-finding evaluations \citep{flournoy2020}. \rev{In this scenario, both mean efficacy and mean toxicity increase with dose, restricting acceptable combinations to the lower-dose region. It tests the allocation and final-selection comparisons under that geometry, using synthetic continuous responses without calibration to a specific clinical program.}

\rev{Our contribution is a deterministic calculation of the one-cohort cKG score and a comparison of allocation objectives under shared decision rules. The accompanying Python package, \href{https://github.com/clairexinzhuwang/dose-combination-bo}{\texttt{dose-combination-bo}}, implements all four acquisitions, with examples for running a trial and comparing rules. The archived study implementation remains the source of the reported operating characteristics.}

\Needspace{8\baselineskip}
\section{Methods}\label{sec:methods}

\subsection{Dose-selection target and response models}

\rev{Let $e(d,z)$ and $g(d,z)$ be mean efficacy and mean toxicity for dose pair $d\in[0,1]^2$ in stratum $z\in\{0,1\}$. Both are continuous functions of dose; larger efficacy and smaller toxicity are preferred.} The upper limit on mean toxicity is $g^\dagger(z)$. \rev{The toxicity score, observation window, and limit are specified simulation inputs.} Assignment and final selection use the $5\times5$ grid $\mathcal G=\{0,0.25,0.50,0.75,1\}^2$. The coordinates are standardized dose levels: \rev{zero denotes the lowest dose level, which need not be placebo or absence of an agent}. Every scenario contains at least one grid combination below the mean-toxicity limit. The target is
\begin{equation}\label{eq:obd_grid}
d^*_{\rm grid}(z)=\arg\max_{d\in\mathcal G}e(d,z)\quad\text{subject to}\quad g(d,z)\le g^\dagger(z).
\end{equation}
This target maximizes efficacy within a toxicity constraint; it does not assign a further penalty to toxicity below that limit or require a minimum efficacy. \rev{The continuous surfaces $e$ and $g$ define simulation truth. GP covariance is also defined over the continuous domain, but assignment and final selection are restricted to $\mathcal G$.} We fit independent Mat\'ern-$5/2$ GPs to efficacy and toxicity. Write $E$ and $G$ for their latent random functions, with posterior means $\mu_e,\mu_g$ and \rev{latent posterior SDs} $\sigma_e,\sigma_g$. Kernel and mean parameters are re-estimated after each observed cohort. \rev{Let $D_t$ contain the observed doses and both outcomes through cohort $t$, including initialization data. For the model and acquisition definitions below, fix a stratum and the current data $D_t$; stratum and time arguments are suppressed unless needed.} Let $\Phi$ be the standard normal distribution function. For $0<\tau<1$, define
\begin{equation}\label{eq:gate}
\begin{aligned}
z_g(d)&=\frac{g^\dagger-\mu_g(d)}{\sigma_g(d)},\qquad
\PF(d)=\Pr\{G(d)\le g^\dagger\mid D_t\}=\Phi\{z_g(d)\},\\
\rev{A_\tau}&\rev{=\{d\in\mathcal G:\PF(d)>\tau\}.}
\end{aligned}
\end{equation}
\rev{$A_\tau$ is the subset of the full dose grid $\mathcal G$ that passes the current posterior criterion.} The probability $\PF(d)$ concerns the unknown \emph{mean} toxicity, not an individual participant's probability of an adverse event. The cutoff $\tau$ is therefore not a target dose-limiting-toxicity rate. For a fixed posterior, increasing $\tau$ makes the requirement more stringent. All rules use the strict inequality in Equation~\eqref{eq:gate}; cEI also uses $\PF(d)$ as a continuous weight.

At the scheduled final analysis, the rule selects \rev{$\hat d=\arg\max_{d\in A_\tau}\mu_e(d)$} if any combination meets the criterion. Otherwise, it selects the grid combination with the largest $z_g$. This \emph{fallback selection does not meet the posterior toxicity criterion}. No minimum-exposure requirement is imposed, so the selected combination may never have been assigned. \rev{Under gradual availability, a trial stopped by the rule below ends without recommending a combination.}

\subsection{\rev{Acquisition functions}}\label{sec:policies}

We compare four ways to score the next cohort's assignment. cEI and cKG use efficacy and toxicity; tMSE and the entropy rule learn only about toxicity. \rev{All four are compared under the same assignment, stopping, and final-selection rules; tMSE and entropy serve as acquisition comparators rather than established trial designs.} \rev{Let $\mathcal G_t\subseteq\mathcal G$ be the combinations currently available}, and let $\mathcal E_t\subseteq\mathcal G_t$ be those permitted for the next assignment. Each acquisition ranks $A_\tau\cap\mathcal E_t$ when that set is nonempty. With only one eligible combination, the acquisition cannot change the assignment. \rev{If no permitted combination meets the criterion, the same highest-$z_g$ fallback is applied within $\mathcal E_t$, subject to the stopping rule below.} Ties are resolved by a fixed grid order.

\rev{For final-selection value, let $C_t=A_\tau\cap\mathcal G_t$ when nonempty; otherwise, $C_t$ is the singleton containing the available combination with the largest $z_g$. Write $V_t=\max_{x\in C_t}\mu_e(x)$ for its best posterior mean efficacy. This value is also the cEI incumbent when the acquisition determines assignment.}

\paragraph{Feasibility-weighted expected improvement (cEI).} \rev{We rank dose combinations by}
\begin{equation}\label{eq:cei}
\alpha_{\rm cEI}(d)=\mathrm{EI}\{\mu_e(d),\sigma_e(d);\rev{V_t}\}\PF(d).
\end{equation}
\rev{EI is the posterior expected positive efficacy gain over $V_t$. The incumbent includes previously visited combinations and follows the posterior-mean convention in archived version~1 of Willard et al., expressed for efficacy maximization} \citep{willardobd}. \rev{Under independent fitted GPs and a fixed incumbent, constrained EI factors into EI times $\PF$} \citep{gardner2014,gelbart2014,letham2019}. \rev{The updated design of Willard et al. instead averages this product over posterior draws of GP parameters and draw-specific constrained optima} \citep{willardobd2026}. \rev{Here cEI uses the same fitted Mat\'ern GPs, assignment restrictions, stopping, and final-selection rules as the other acquisitions, so the comparison concerns allocation objectives under our common protocol.} It favors combinations with larger expected efficacy improvement and stronger posterior support for meeting the toxicity limit. It does not explicitly calculate how a new cohort would change the final selection.

\paragraph{One-cohort lookahead (cKG).} \rev{For each eligible dose pair, cKG averages the posterior mean efficacy of the selection that would follow a new cohort, accounting for its possible efficacy and toxicity outcomes. The calculation uses only currently available combinations. Let $\bar Y_e(d)$ and $\bar Y_g(d)$ be the hypothetical cohort means at $d$. Averaging over both outcomes gives the score}
\begin{equation}\label{eq:kg}
\revisioncolor
\alpha_{\mathrm{cKG}}(d)=\mathbb E\{V_t^+(d)\mid D_t\}-V_t.
\end{equation}
\rev{Each hypothetical cohort updates both GPs with the current hyperparameters held fixed; a superscript $+$ denotes the updated quantities. Let $A_\tau^+=\{x\in\mathcal G:\PF^+(x)>\tau\}$. The updated selection set is $C_t^+=A_\tau^+\cap\mathcal G_t$ when this intersection is nonempty; otherwise it is the singleton containing the available combination with the largest $z_g^+$. We set $V_t^+(d)=\max_{x\in C_t^+}\mu_e^+(x)$ and assign the next cohort to the eligible combination with the largest cKG score.} \rev{When $A_\tau\cap\mathcal E_t$ is nonempty, maximizing cKG maximizes the expected selection value after one cohort under the fitted models.} Under the independent Gaussian models, toxicity updates partition the calculation into intervals with a fixed selection set; the efficacy expectation can be integrated analytically within each interval (Appendix~\ref{sec:ckg_evaluator}). \rev{Appendix~\ref{sec:theory} gives the score decomposition.} Later cohorts, future dose availability, stopping, and the hyperparameter refit following an observed cohort are not included in the lookahead.

\paragraph{Targeted mean squared error (tMSE).} With equal candidate weights, the zero-bandwidth noisy-level-set form of the criterion introduced by \citet{picheny2010} and studied by \citet{lyu2021} ranks combinations by
\begin{equation}\label{eq:tmse}
\rev{\alpha_{\rm tMSE}}(d)=\sigma_g(d)\exp\{-z_g(d)^2/2\}
\end{equation}
up to a positive constant common to all candidates. tMSE favors combinations with uncertain toxicity and \rev{a posterior mean toxicity near the limit}, but assignment remains subject to $A_\tau\cap\mathcal E_t$. It uses neither efficacy nor the effect of a new cohort on the final dose choice.

\paragraph{\rev{Learning toxicity across the grid (entropy).}} \rev{Let $h(p)=-p\log p-(1-p)\log(1-p)$ be binary entropy, with $0\log0=0$. The entropy score is}
\begin{equation}\label{eq:sur}
\revisioncolor
\alpha_{\rm ent}(d)=\sum_{x\in\mathcal G_t}\left[
h\{\PF(x)\}-\mathbb E\{h(\PF^+(x))\mid D_t\}\right].
\end{equation}
The expectation averages over possible toxicity cohort means $\bar Y_g(d)$; $\PF^+$ uses each resulting update, with fitted parameters and $\mathcal G_t$ held fixed. The score sums the expected reduction in uncertainty about whether each available combination meets the mean-toxicity limit. Uncertainty is calculated separately for each combination, rather than for the acceptable set as a whole. The entropy-based design of \citet{domenicano2019} learns which dose is optimal. Here, we learn only toxicity, drawing on methods that identify where a response surface falls below a threshold \citep{bect2012,chevalier2014,marques2018}. Numerical details and checks are in Appendix~\ref{sec:numerical_impl}.

\subsection{\rev{Trial designs and outcome measures}}\label{sec:protocol}

\rev{We compare full-grid and gradual dose availability. In the full-grid design, each stratum contributes 40 response records: four initialization observations generated at stratified continuous dose locations, followed by 18 two-participant cohorts assigned on $\mathcal G$. There is no early stopping. Allocation summaries use the 36 post-initialization assignments. The four initial observations are model inputs; obtaining them in a clinical trial would require a separate initialization protocol.}

In the gradual-availability analysis, enrollment is capped at 40 participants per stratum. \rev{The first cohort of two participants starts at $(0,0)$}, after which a fixed schedule makes more combinations available. These initial on-grid assignments count toward enrollment and allocation outcomes. The schedule is not response-guided escalation. Before the full grid is available, previously assigned combinations are excluded while any available unassigned combination remains. This restriction can prevent returning to an observed, tolerated combination. The assignment fallback is used on the first two consecutive occasions when no permitted combination passes the criterion; \rev{a third consecutive occasion stops the trial without selection}. At maximum enrollment, final selection uses the full grid. \rev{Because availability, eligibility, stopping, and enrollment change together, this analysis cannot isolate the effect of escalation.} Appendix~\ref{sec:phase} gives the schedule.

\rev{We report true mean efficacy at final selection and how often that selection exceeds the true mean-toxicity limit, retaining fallback selections. For gradual availability, efficacy is summarized only among trials that make a selection. The above-limit selection percentage is reported both among all initiated trials and among selecting trials, alongside the stopping rate. Stopped trials contribute no above-limit selection to the all-trial percentage; no efficacy value is assigned to them.}

\rev{The main allocation outcome is the percentage of the 36 post-initialization participants assigned above the limit in full-grid runs. Under gradual availability, where enrollment varies, we report the mean number $N_{\rm above}$ assigned above the limit per initiated trial, including the two initial participants, together with mean enrollment. Both outcomes concern assignments to combinations with above-limit true mean toxicity, not observed adverse events. Appendix~\ref{sec:outcomes} gives the formal definitions.}

\subsection{\rev{Simulation scenarios}}\label{sec:analysis}

We use four synthetic scenarios: OSA-derived \citep{willardobd}, logistic, MARIPOSA-motivated \citep{schweitzer2023}, and Gaussian bump. \rev{The Gaussian-bump scenario has nonmonotone mean toxicity with substantial changes between adjacent grid points (Appendix~\ref{sec:demo}).} \rev{Let $M$ denote the number of independent simulation replicates. The first and last scenarios use $M=200$; the others use $M=100$.} Within a replicate, rules share random-number streams for initialization and outcome errors, although adaptive assignments can diverge. \rev{Each scenario is evaluated in two strata and at $\tau\in\{0.5,0.6,0.7,0.8,0.9\}$, with 0.7 as the reference.} Gradual availability is evaluated for OSA at 0.7 and 0.9. Strata have separate simulated trials; \rev{shared accrual and cross-stratum borrowing are not evaluated}. \rev{Efficacy and toxicity outcomes are continuous with Gaussian errors.} Both outcomes from each cohort are assumed available before the next assignment; accrual time, delayed outcomes, and missingness are not simulated.

\rev{For OSA, efficacy is simulated AHI4 reduction in events/hour on the published synthetic response surface \citep{willardobd}; larger values represent greater reduction. The OSA strata represent mild/moderate ($z=0$) and severe ($z=1$) disease; toxicity uses the synthetic natural-log scale $\log(B+0.5)$, where $B$ is a weighted adverse-event burden score, with mean-score limits of 1.5 and 2.0, respectively. The other three scenarios use their own synthetic efficacy scales. Differences between percentages are reported in percentage points (pp).}

We summarize paired replicate-level contrasts by their mean, Monte Carlo standard error (MCSE), and two-sided 95\% $t$ interval. \rev{Each surface-level summary weights two strata and five $\tau$ settings equally within replicate. We report surfaces separately; each average describes the settings evaluated on that surface, rather than a clinical population or a single trial design.} \rev{The main comparisons are exploratory, with pointwise, unadjusted intervals.} Monte Carlo intervals quantify simulation error under the specified data-generating models, not uncertainty about clinical treatment effects.

\subsection{\rev{Design of the sensitivity analysis}}\label{sec:calibrationaudit}

A nested OSA analysis compares all four rules at $\tau=0.7,0.9$ without early stopping. Each path is analyzed after 20, 40, and 80 total response records, corresponding to 16, 36, and 76 post-initialization assignments. The analysis model assumes a toxicity-outcome SD equal to half, the same as, or twice the data-generating SD, denoted $c_\sigma\in\{0.5,1,2\}$. The true surfaces, true outcome SDs, and efficacy model are unchanged. Allocation is summarized as a percentage of post-initialization assignments. Simultaneous 95\% max-$t$ Monte Carlo intervals cover the 54 displayed contrasts. \rev{This analysis is limited to the assumed toxicity-outcome SD; it does not calibrate clinical enrollment or assess other model errors.}

Replication followed a prespecified Monte Carlo precision rule; Appendix~\ref{sec:noiserobust} gives the rule and its cap \citep{morris2019,koehler2009,burton2006}.

\subsection{\rev{Software and computation}}

\rev{In a fixed-data CPU benchmark of the installed pre-rename software candidate 0.2.1rc4, cKG scoring on the 25-point dose grid took 0.105--0.114 seconds; fitting both response GPs took 0.136--0.147 seconds. The benchmark used three fixed OSA datasets with 4, 20, and 40 observations, rather than adaptive trial paths (Appendix~\ref{sec:runtime}).}

\section{Results}\label{sec:results}

\subsection{\rev{Efficacy and toxicity in the OSA-derived scenario}}

\rev{Table~\ref{tab:osa_absolute_oc} summarizes efficacy and toxicity for all four rules.} Relative to cEI, cKG increased final-selection efficacy by \rev{0.739 events/hour} (95\% Monte Carlo interval, 0.504 to 0.974), above-limit final selection by 8.30 \rev{pp} (5.74 to 10.86), and above-limit allocation by 2.82 \rev{pp} (1.53 to 4.11). \rev{The efficacy gain came with more above-limit assignments and final selections, so it did not establish an overall benefit--risk advantage.} This summary weights five cutoff settings equally. \rev{Table~\ref{tab:cei_joint} reports all three outcomes separately at each cutoff.} \rev{Appendix~\ref{sec:nearoptimal} reports acceptable selection within an efficacy margin of the grid optimum.}

\begin{table}[H]
\revisioncolor
\centering
\footnotesize
\setlength{\tabcolsep}{4pt}
\caption{Absolute operating characteristics in the primary OSA-derived analysis under full-grid availability.}
\label{tab:osa_absolute_oc}
\begin{tabular}{@{}lccc@{}}
\toprule
Rule & \shortstack{\rev{AHI4 reduction}\\\rev{at final selection (events/hour)}} & \shortstack{Above-limit final\\selection (\%)} & \shortstack{Above-limit\\allocation (\%)} \\
\midrule
cKG & 7.46 (0.15) & 18.80 (1.19) & 17.92 (0.61) \\
tMSE & 7.36 (0.14) & 11.85 (1.09) & 27.08 (0.79) \\
Entropy rule & 8.02 (0.15) & 18.65 (1.50) & 22.42 (0.78) \\
cEI & 6.72 (0.15) & 10.50 (1.14) & 15.10 (0.84) \\
\bottomrule
\end{tabular}
\vspace{3pt}
\begin{minipage}{0.98\linewidth}
\scriptsize\raggedright
\textit{Note.} \rev{Entries are descriptive means over $M=200$ replicates, weighting the two strata and five $\tau$ values equally. Parentheses give Monte Carlo standard errors (MCSEs), not between-trial standard deviations.} Above limit refers to true mean toxicity above the prespecified limit. Allocation uses the 36 post-initialization participants and excludes the four initialization records. All trials made a final selection (100\%), including prespecified fallback selections. \rev{Efficacy is simulated AHI4 reduction in events/hour.} Table~\ref{tab:aggregate} contrasts were calculated before rounding.
\end{minipage}
\end{table}

\begin{table}[H]
\centering
\footnotesize
\setlength{\tabcolsep}{3pt}
\caption{Joint efficacy--toxicity contrasts for cKG versus cEI in the OSA-derived scenario.}
\label{tab:cei_joint}
\begin{tabular}{@{}cccc@{}}
\toprule
$\tau$ & \shortstack{Final-selection\\efficacy difference} & \shortstack{Above-limit final\\selection (pp)} & \shortstack{Above-limit\\allocation (pp)} \\
\midrule
0.5 & 1.19 [0.76, 1.63] & 13.00 [6.82, 19.18] & 6.50 [4.38, 8.62] \\
0.6 & 1.03 [0.62, 1.45] & 10.50 [5.43, 15.57] & 4.07 [2.20, 5.94] \\
0.7 & 0.91 [0.51, 1.31] & 9.75 [5.18, 14.32] & 2.33 [0.69, 3.98] \\
0.8 & 0.39 [0.04, 0.74] & 5.50 [1.95, 9.05] & 1.33 [0.01, 2.65] \\
0.9 & 0.17 [$-0.13$, 0.47] & 2.75 [0.12, 5.38] & $-0.14$ [$-1.33$, 1.06] \\
\bottomrule
\end{tabular}
\vspace{3pt}
\begin{minipage}{0.98\linewidth}
\scriptsize\raggedright
\textit{Note.} Entries are cKG-minus-cEI differences [pointwise unadjusted 95\% Monte Carlo intervals], averaging the two strata within each of $M=200$ paired replicates. Positive efficacy differences favor cKG; \rev{positive differences in above-limit selection or allocation favor cEI}. \rev{Efficacy differences are in events/hour.} Allocation uses 36 post-initialization assignments. Final-selection results retain fallback selections.
\end{minipage}
\end{table}

\begin{figure}[t]
\centering
\includegraphics[width=0.98\linewidth]{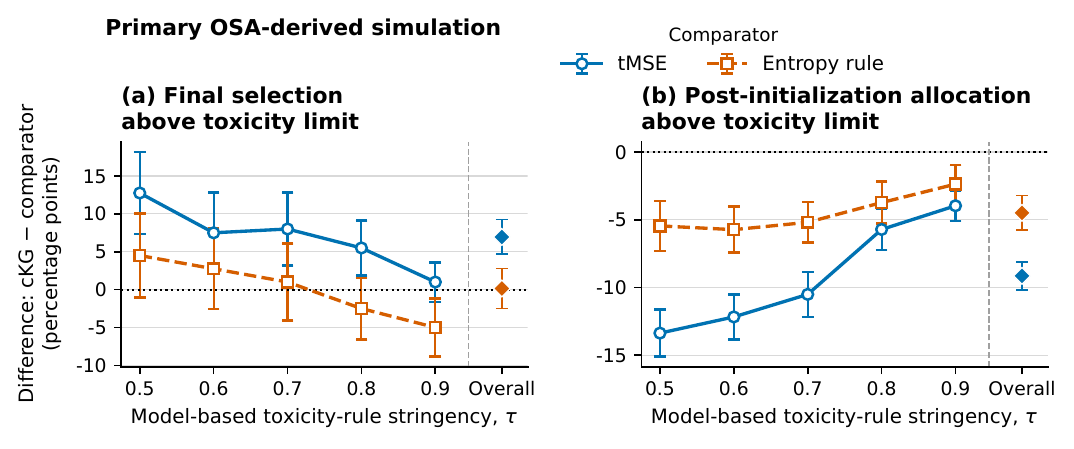}
\caption{cKG-minus-comparator differences in the primary OSA-derived simulation ($M=200$): (a) above-limit final selection and (b) above-limit post-initialization allocation. Points average the two strata; ``Overall'' is a descriptive average over the five $\tau$ values, not a single trial design. Bars are pointwise unadjusted 95\% Monte Carlo intervals.}
\label{fig:family_profile}
\end{figure}

\rev{In OSA, lower above-limit allocation did not imply fewer above-limit final selections.} Relative to tMSE, cKG reduced above-limit allocation by 9.15 \rev{pp} (95\% Monte Carlo interval, 8.12 to 10.19), but increased above-limit final selection by 6.95 \rev{pp} (4.70 to 9.20). Relative to entropy, the allocation difference favored cKG by 4.50 \rev{pp} (3.24 to 5.76); the final-selection difference was 0.15 \rev{pp} ($-2.48$ to 2.78). That interval does not establish equivalence. Entropy also had higher mean efficacy at final selection than cKG, \rev{8.02 versus 7.46 events/hour} (Table~\ref{tab:osa_absolute_oc}). Table~\ref{tab:aggregate} and Figure~\ref{fig:family_profile} give the paired contrasts.

\subsection{Efficacy of the final selection: cKG versus cEI}

\begin{figure}[htbp]
\centering
\includegraphics[width=0.92\linewidth]{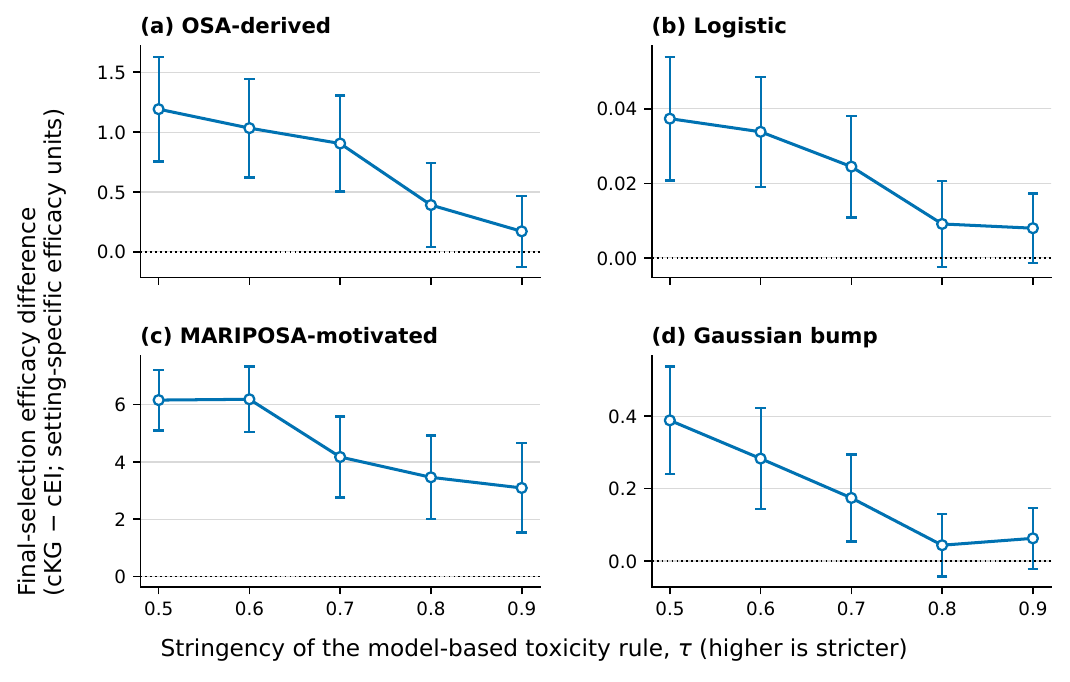}
\caption{cKG-minus-cEI differences in true mean efficacy at final selection under full-grid availability. Points average the two strata at each $\tau$; bars are pointwise unadjusted 95\% Monte Carlo intervals. Positive values indicate higher true mean efficacy under cKG. Efficacy scales differ by setting. OSA-derived and Gaussian-bump settings use $M=200$ replicates; the others use $M=100$.}
\label{fig:cei_efficacy}
\end{figure}

\rev{cKG had higher estimated mean efficacy at final selection than cEI at every cutoff on all four surfaces} (Figure~\ref{fig:cei_efficacy}; Appendix Table~\ref{tab:ckg_cei_efficacy_tau}). Pointwise Monte Carlo intervals excluded zero at $\tau=0.5,0.6,0.7$ on every surface. Differences were smaller at 0.8 and 0.9; some intervals then included zero. Efficacy is not pooled across surfaces because its scale differs. \rev{In the Gaussian-bump scenario, fallback occurred in 17.25--41.75\% of cKG trials and 17.25--41.25\% of cEI trials, depending on $\tau$. The pooled cKG--cEI efficacy difference was 0.19 overall, compared with 0.24 among criterion-passing selections and 0.06 among fallback selections (Appendix~\ref{sec:selection_partition}). These subgroup comparisons involve different trials under the two rules and do not show how either rule would perform without fallback.}

\rev{Fallback accounted for little of the excess above-limit final selections under cKG in OSA} (Appendix Table~\ref{tab:selection_partition}). Criterion-passing selections were truly above limit in 18.55\% of all cKG trials and 10.15\% of cEI trials; above-limit fallback selections accounted for only 0.25\% and 0.35\%. The probability of a selection that both passed the criterion and was truly acceptable was 80.55\% under cKG versus 88.30\% under cEI, a difference of $-7.75$ \rev{pp} (95\% Monte Carlo interval, $-10.38$ to $-5.12$). On the MARIPOSA-motivated surface neither policy used a terminal fallback, yet their above-limit selection rates were 20.40\% and 0.20\%. \rev{Passing the posterior criterion therefore did not ensure that true mean toxicity was below the limit.}

\subsection{Patterns across synthetic response surfaces}

In each surface-level summary, cKG made more above-limit final selections than tMSE, by 4.25--12.10 \rev{pp}, while allocating less often above the limit, by 2.36--11.37 \rev{pp}. Pointwise Monte Carlo intervals excluded zero for both outcomes in all four summaries. The Gaussian-bump scenario has a different, nonmonotone geometry (Appendix Figure~\ref{fig:simsurf}), but several features change together; these comparisons do not identify which feature explains the pattern.

Comparisons with entropy were less consistent. On the logistic surface, cKG both selected above-limit combinations more often and allocated above the limit less often, with intervals excluding zero. On the OSA-derived and MARIPOSA-motivated surfaces, only the allocation intervals excluded zero; on the Gaussian-bump surface, neither interval excluded zero. Appendix Figure~\ref{fig:family_quadrant_cellwise} shows the complete surface-by-$\tau$ profiles.

\rev{The efficacy comparisons also varied by surface. Final-selection efficacy favored cKG over tMSE on the logistic and Gaussian-bump surfaces. Compared with entropy, efficacy favored cKG on the logistic surface and entropy on OSA; the remaining pointwise intervals included zero (Appendix Table~\ref{tab:efficacy_comparators}).}

\begin{table}[H]
\revisioncolor
\centering
\footnotesize
\setlength{\tabcolsep}{3pt}
\caption{cKG-minus-comparator differences in above-limit final selection and participant allocation by simulation scenario under full-grid availability.}
\label{tab:aggregate}
\begin{tabular}{@{}>{\raggedright\arraybackslash}p{0.23\linewidth}>{\raggedright\arraybackslash}p{0.13\linewidth}>{\raggedright\arraybackslash}p{0.27\linewidth}>{\raggedright\arraybackslash}p{0.27\linewidth}@{}}
\toprule
Setting & Comparator & \shortstack{Above-limit final\\selection (pp)} & \shortstack{Above-limit\\allocation (pp)} \\
\midrule
OSA-derived & tMSE & 6.95 (1.14) [4.70, 9.20] & $-9.15$ (0.52) [$-10.19$, $-8.12$] \\
OSA-derived & Entropy rule & 0.15 (1.33) [$-2.48$, 2.78] & $-4.50$ (0.64) [$-5.76$, $-3.24$] \\
Logistic & tMSE & 12.10 (1.68) [8.76, 15.44] & $-5.32$ (0.60) [$-6.52$, $-4.13$] \\
Logistic & Entropy rule & 11.90 (1.76) [8.40, 15.40] & $-5.07$ (0.61) [$-6.27$, $-3.86$] \\
MARIPOSA-motivated & tMSE & 5.40 (2.68) [0.08, 10.72] & $-11.37$ (0.99) [$-13.32$, $-9.41$] \\
MARIPOSA-motivated & Entropy rule & 3.60 (2.92) [$-2.19$, 9.39] & $-7.78$ (0.98) [$-9.72$, $-5.83$] \\
Gaussian bump & tMSE & 4.25 (1.06) [2.16, 6.34] & $-2.36$ (0.58) [$-3.50$, $-1.22$] \\
Gaussian bump & Entropy rule & 1.60 (1.23) [$-0.82$, 4.02] & $-0.10$ (0.74) [$-1.57$, 1.36] \\
\bottomrule
\end{tabular}
\vspace{3pt}
\begin{minipage}{0.98\linewidth}
\scriptsize\raggedright
\textit{Note.} Entries are differences in pp shown as estimate (MCSE) [pointwise unadjusted 95\% Monte Carlo interval]. Above limit refers to true mean toxicity above the prespecified limit. Allocation uses the 36 post-initialization participants and excludes the four initialization records. Within each scenario, the two strata and five $\tau$ values are weighted equally; these averages are descriptive. OSA-derived and Gaussian-bump rows use $M=200$ replicates; logistic and MARIPOSA-motivated rows use $M=100$. Positive selection differences and negative allocation differences mean more above-limit selections and fewer above-limit assignments under cKG, respectively. Final-selection results include prespecified fallback selections.
\end{minipage}
\end{table}

\subsection{Gradual dose availability and stopping}

Table~\ref{tab:gradual} reports gradual-availability results for OSA. At $\tau=0.7$, cKG made above-limit final selections 8.75 \rev{pp} more often than tMSE (95\% Monte Carlo interval, 3.89 to 13.61), \rev{while reducing mean $N_{\rm above}$ by 4.50 participants per initiated trial, including stopped trials (3.97 to 5.03 fewer).} At $\tau=0.9$, the corresponding differences were 2.50 \rev{pp} (0.32 to 4.68) and 1.81 fewer participants (1.42 to 2.19 fewer). \rev{cKG also had lower mean $N_{\rm above}$ than entropy at both $\tau=0.7$ and $0.9$, with intervals excluding zero.} \rev{These are counts per initiated trial, not assignment probabilities: cKG stopped more often than the other rules and consequently enrolled fewer participants. Among trials that made a final selection, cKG had lower mean efficacy than tMSE and entropy at both cutoffs, and higher mean efficacy than cEI; these conditional summaries involve different subsets of trials.}

\begin{table}[H]
\revisioncolor
\centering
\footnotesize
\setlength{\tabcolsep}{4pt}
\caption{OSA-derived operating characteristics under gradual dose availability.}
\label{tab:gradual}
\begin{tabular}{@{}clcccccc@{}}
\toprule
& & & \multicolumn{2}{c}{Above-limit final selection (\%)} & & & \\
\cmidrule(lr){4-5}
$\tau$ & Rule & \shortstack{Final-selection\\efficacy} & \shortstack{All initiated\\trials} & \shortstack{With selection\\($n_{\rm sel}$)} & \shortstack{Stopped\\(\%)} & \shortstack{Mean\\enrollment} & \shortstack{Mean\\$N_{\rm above}$} \\
\midrule
0.7 & cKG & 7.26 & 18.50 & 19.07 (388) & 3.00 & 39.14 & 6.47 \\
0.7 & tMSE & 7.70 & 9.75 & 9.95 (392) & 2.00 & 39.42 & 10.97 \\
0.7 & Entropy rule & 8.10 & 17.50 & 17.54 (399) & 0.25 & 39.92 & 10.07 \\
0.7 & cEI & 6.40 & 10.25 & 10.35 (396) & 1.00 & 39.71 & 3.81 \\
0.9 & cKG & 5.42 & 4.50 & 5.41 (333) & 16.75 & 35.20 & 4.44 \\
0.9 & tMSE & 5.70 & 2.00 & 2.29 (350) & 12.50 & 36.45 & 6.24 \\
0.9 & Entropy rule & 6.07 & 7.50 & 8.11 (370) & 7.50 & 37.86 & 5.78 \\
0.9 & cEI & 5.19 & 2.75 & 3.00 (367) & 8.25 & 37.63 & 3.28 \\
\bottomrule
\end{tabular}
\vspace{3pt}
\begin{minipage}{0.98\linewidth}
\scriptsize\raggedright
\textit{Note.} Each rule--$\tau$ row represents 400 initiated stratum-specific trials ($M=200$ replicates in two strata); larger $\tau$ denotes a more stringent toxicity rule. Above limit refers to true mean toxicity above the prespecified limit. Stopped trials contribute zero to the all-trial percentage; parentheses in the conditional column give $n_{\rm sel}$. \rev{Efficacy is mean simulated AHI4 reduction (events/hour) among the $n_{\rm sel}$ trials that made a selection}; no efficacy value is assigned to stopped trials. \rev{The set of trials contributing to each efficacy mean differs between rules.} Mean enrollment and mean $N_{\rm above}$ use all initiated trials. $N_{\rm above}$ includes the two initial on-grid assignments and counts assigned participants, not observed toxicity events.
\end{minipage}
\end{table}

\Needspace{8\baselineskip}
\subsection{Sensitivity to enrollment and assumed toxicity variability}

\rev{The nested analysis used $M=500$ replicates, evaluated after 20, 40, or 80 response records with assumed-to-generated toxicity-SD ratios $c_\sigma\in\{0.5,1,2\}$. Of the 54 toxicity contrasts, 51 met the prespecified precision criterion.} Appendix~\ref{sec:noiserobust} describes the procedure and exceptions.

Compared with tMSE, cKG reduced above-limit allocation by 3.08--8.65 \rev{pp} across the snapshots and assumed SDs; all simultaneous intervals excluded zero. Above-limit final selection was 1.90--9.30 \rev{pp} more frequent, with intervals excluding zero except at the earliest snapshot when toxicity variability was understated. Neither contrast followed a common monotone pattern across the three SD assumptions.

Results against entropy changed with the assumed toxicity SD. When the SD was understated, cKG allocated less often above the limit at all snapshots and selected above-limit combinations more often at the last two. When the SD was overstated, the allocation contrast reversed at the first two snapshots, and cKG selected above-limit combinations less often at the last two. Under the matching SD, cKG allocated above the limit more often at the first snapshot and less often thereafter; the final-selection interval excluded zero only at the last snapshot. Compared with cEI, cKG made above-limit final selections more often in every setting, with simultaneous intervals excluding zero. \rev{The ranking depended on the assumed toxicity SD. Across the nine SD-by-snapshot summaries, cKG had higher mean final-selection efficacy than cEI and lower mean efficacy than entropy; its efficacy relative to tMSE depended on the setting (Appendix Table~\ref{tab:nested_efficacy}). These efficacy summaries are descriptive and separate from the 54 toxicity contrasts.}

\section{Discussion}\label{sec:discussion}

\rev{The deterministic evaluator calculates a one-cohort cKG score under independent Gaussian response models, with fitted hyperparameters held fixed during each hypothetical update. In the full-grid simulations, cKG had higher estimated mean final-selection efficacy than cEI at every cutoff on all four surfaces, with smaller differences under stricter criteria. Averaged over the OSA strata and cutoffs, this gain accompanied more above-limit assignments and final selections. Compared with tMSE, cKG assigned fewer participants to above-limit combinations but selected them more often at trial completion. These outcomes describe distinct trade-offs rather than a single ranking of the rules.}

\rev{Entropy had higher final-selection efficacy than cKG in the primary OSA summary. Although it does not target efficacy directly, reducing uncertainty about toxicity may allow more high-efficacy combinations to qualify for final selection. cKG also learns toxicity, but assigns the next cohort according to its expected effect on selection value under the current models. Computing that score exactly does not guarantee better performance over the whole trial.}

\rev{For trial statisticians, the starting point is the final decision: the toxicity criterion, any minimum efficacy requirement, the evidence needed to recommend a combination, and the option to recommend none. Within a finite-grid BO design using the independent fitted GPs and recommendation rule studied here, cKG can replace cEI for cohort allocation while eligibility, stopping, and final selection remain fixed. The complete protocol should then be evaluated, reporting efficacy together with above-limit allocation, final-selection toxicity, fallback use, and stopping. The \texttt{dose-combination-bo} package supports these comparisons. Our cEI shares the acquisition form used by Willard et al., but the comparison does not reproduce their complete design \citep{willardobd,willardobd2026}. In particular, their updated fully Bayesian fitting and optimal-dose entropy stopping differ from our fitted-model comparison and toxicity-classification entropy acquisition.}

\Needspace{5\baselineskip}
\rev{Several features limit clinical interpretation. Posterior probability cutoffs were not clinically calibrated. The selection rule imposes no additional toxicity penalty below the limit and can recommend an untried combination or a fallback that fails the criterion. Gaussian-bump results include many fallbacks; OSA and MARIPOSA-motivated results also show that criterion-passing selections can exceed the true mean-toxicity limit. The primary design uses four initialization observations, while gradual availability starts on-grid and follows a fixed opening schedule. Both assume that cohort outcomes are immediately available. Neither addresses delayed outcomes, dropout, shared accrual, permanent dose elimination, or independent safety monitoring. Every scenario contains an acceptable grid dose; no minimum clinically important efficacy or all-inactive scenario is specified. Independent GPs omit efficacy--toxicity dependence, and plug-in fitting omits hyperparameter uncertainty. The primary simulations also use the data-generating outcome variances. The sensitivity analysis varies only the assumed toxicity-outcome SD, not kernel choice, dose scaling, or prior-data conflict. A clinical application would require justified outcome definitions, observation windows and toxicity limits, separate handling of serious adverse events and discontinuation, and case-specific validation of the complete decision rules. Published arm summaries cannot supply the required individual-level outcome distributions.}

\section{Conclusions}

The proposed cKG rule calculates how one additional cohort would change the value of the final dose-selection rule under the fitted response models. \rev{Under full-grid availability, cKG had higher estimated mean final-selection efficacy than cEI at every tested cutoff on all four surfaces. Pointwise Monte Carlo intervals excluded zero at $\tau=0.5,0.6,0.7$, but not in every stricter-cutoff comparison. Averaged over the OSA strata and cutoffs, cKG made more above-limit assignments and final selections than cEI, and fewer above-limit assignments but more above-limit final selections than tMSE.} \rev{Allocation and final selection should therefore be evaluated separately, with efficacy reported alongside toxicity, fallback use, and stopping.} \rev{The accompanying software provides these allocation objectives for evaluation under explicitly specified toxicity criteria and final-selection rules.}

\section*{Conflict of interest}
The authors declare no conflict of interest.

\section*{Funding}
No specific funding was received for this work.

\section*{Ethics approval}
Not applicable; this study used simulated data and published arm-level summaries only.

\section*{Patient consent}
Not applicable.

\section*{Clinical trial registration}
Not applicable.

\section*{Data availability statement}
\rev{No patient-level clinical data were used. The maintained software, frozen controller reference, selected simulation records, analysis scripts, and summary outputs are held at \url{https://github.com/clairexinzhuwang/dose-combination-bo}. The included records support recalculation of full-grid and gradual-availability summaries and the post hoc selection-margin analysis. The complete original study archive, including the enrollment--noise path shards required to replay calibration snapshots, is retained separately and is not included in this repository. Public access to that complete archive and a persistent archive identifier remain to be established for this candidate.}

\section*{Use of generative AI}
Generative AI tools, including OpenAI Codex, assisted language editing, code review, \rev{preparation and execution of verification and post-processing scripts, and software packaging}. \rev{AI-assisted verification and post-processing scripts were run separately from the original simulation experiments.} The authors remain responsible for reviewing the generated material, analysis, interpretation, references, and final submitted text.

\section*{Author contributions}
X.W. conceived the study, adapted the method, developed its evaluator, implemented and ran the simulations, and drafted the manuscript. T.L. supervised the work and reviewed and edited the manuscript.

\begingroup
\small
\setlength{\bibsep}{2pt}
\revisioncolor
\renewcommand{\bibnumfmt}[1]{#1.}

\endgroup

\clearpage
\appendix

\section{Implementation and simulation details}\label{sec:phase}

\subsection{Response models and assignment implementation}\label{sec:acqdetails}

Conditionally on dose and stratum, $y_e=e(d,z)+\varepsilon_e$ and $y_g=g(d,z)+\varepsilon_g$, with independent mean-zero Gaussian errors at the within-combination SDs specified below. The simulations do not generate binary dose-limiting toxicity outcomes. The model-based quantity $\PF(d)=\Pr\{G(d)\le g^\dagger\mid D_t\}=\Phi\{z_g(d)\}$, where $\Phi$ is the standard normal distribution function, concerns latent mean toxicity, not an individual participant's toxicity probability. The implementation tests $z_g(d)>\Phi^{-1}(\tau)$, which is equivalent to $\PF(d)>\tau$.

{\revisioncolor
Let $\mathcal E_t\subseteq\mathcal G_t$ be the subset eligible for assignment under the simulated protocol rules. Whenever $A_\tau\cap\mathcal E_t$ is nonempty, each acquisition is maximized over that set. Otherwise, the acquisition score does not determine the assignment, and the fallback selects
\begin{equation}\label{eq:si_emptygate}
d_t^{\rm fallback}=\arg\max_{d\in\mathcal E_t}z_g(d).
\end{equation}
Ties are broken by the fixed grid order. Because the fallback combination does not meet the model-based toxicity criterion, Equation~\eqref{eq:si_emptygate} is only a simulation-continuation rule; clinical use would require separate safety rules.

}

\paragraph{GP fitting.} Each response has a constant-mean GP with a scaled Mat\'ern-$5/2$ kernel and a separate lengthscale for each dose axis. After an observed cohort, kernel and mean parameters are fitted by 120 Adam steps on the marginal likelihood with learning rate 0.1. This covariance permits smooth variation over the dose grid but imposes neither monotonicity nor a pharmacologic interaction model. The choice is shared across acquisitions, not selected as clinically optimal. The primary analysis uses the data-generating within-combination outcome variances; these are not estimated from the small simulated trial. $\PF(d)$ uses latent-function variance. \rev{Let $\sigma_{ye}$ and $\sigma_{yg}$ denote the participant-outcome SDs. For a cohort of two, independent outcome errors contribute $\sigma_{ye}^2/2$ and $\sigma_{yg}^2/2$ to the efficacy and toxicity cohort-mean variances, respectively. These differ from the latent posterior variances $\sigma_e^2(d)$ and $\sigma_g^2(d)$.} Hyperparameters remain fixed during each hypothetical update and are re-estimated after the actual simulated cohort.

\paragraph{Independent response models.} \rev{All rules use separate efficacy and toxicity GPs. We did not study multi-output borrowing, which could change posterior uncertainty and allocation; independence does not imply equivalence to a joint model} \citep{bonilla2007,alvarez2012}.

\subsection{\rev{What the allocation scores measure}}\label{sec:theory}

\rev{The four allocation scores value different information about dose response. For cKG, we can separate the effects of efficacy learning and changes in which combinations meet the toxicity criterion.} For a candidate next-cohort combination $q$, let $C_t$ be the current value set defined in Section~\ref{sec:policies}, and let $C_t^+$ be the corresponding set after a hypothetical cohort. Adding and subtracting the updated efficacy maximum over $C_t$ gives
\begin{align}\label{eq:decomp_main}
\alpha_{\mathrm{cKG}}(q)
={}&\left[\mathbb E\!\left\{\max_{d\in C_t}\mu_e^+(d)\right\}-\max_{d\in C_t}\mu_e(d)\right] \\
&+\mathbb E\!\left\{\max_{d\in C_t^+}\mu_e^+(d)-\max_{d\in C_t}\mu_e^+(d)\right\}.
\end{align}
The first term is nonnegative by the posterior martingale property and Jensen's inequality, conditional on the fitted GP and fixed hyperparameters. \rev{It measures efficacy learning with the current selection set fixed.} The second term accounts for changes to that set after the toxicity update and can have either sign. \rev{A negative cKG score can therefore reflect a decrease in the value of this constrained selection rule. It need not indicate numerical error or imply that information has negative value in an unrestricted decision problem.}

cEI weights local expected efficacy improvement by $\PF(q)$. tMSE depends only on $z_g(q)$ and $\sigma_g(q)$, so candidates with the same local toxicity summaries receive the same score even if they inform other combinations differently. The entropy score sums the mutual information between each indicator $\mathbf 1\{G(x)\le g^\dagger\}$ and the new toxicity observation at $q$, conditional on $D_t$. \rev{It values classification information equally at each currently available grid combination and has no efficacy term.} cKG instead carries both response updates through the final-selection rule.

Increasing $\tau$ from $\tau_1$ to $\tau_2$ removes, for a fixed posterior, combinations with $\Phi^{-1}(\tau_1)<z_g(d)\le\Phi^{-1}(\tau_2)$. For cKG, this also changes which hypothetical toxicity updates alter the selection set. For cEI, the local weight remains $\Phi\{z_g(q)\}$ among eligible candidates. Neither formula implies a fixed ordering of cKG and cEI across cutoffs. \rev{Changing $\tau$ also changes the data collected over a trial; nesting the selection sets for a fixed posterior does not guarantee an ordering of realized allocation or selection errors.}

\subsubsection{Deterministic evaluation of cKG on a finite dose set}\label{sec:ckg_evaluator}

\rev{Fix a query $q$ and index the combinations in $\mathcal G_t$ by $i$, with dose $d_i$. Write $\mu_{e,i}=\mu_e(d_i)$ and $\mu_{g,i}=\mu_g(d_i)$. Conditional on the current GP fit, the updated posterior means have the affine form below, using independent $Z_e,Z_g\sim\mathcal N(0,1)$:}
\begin{equation}\label{eq:ckg_affine_updates}
 \mu_{e,i}^{+}=\mu_{e,i}+b_{e,i}Z_e,\qquad
 \mu_{g,i}^{+}=\mu_{g,i}+b_{g,i}Z_g,
\end{equation}
\rev{Here $b_{e,i}=\operatorname{Cov}\{E(d_i),\bar Y_e(q)\mid D_t\}/\sqrt{\operatorname{Var}\{\bar Y_e(q)\mid D_t\}}$, with $b_{g,i}$ defined analogously for toxicity.} For numerical stability, the implementation uses the floored post-update latent toxicity variance
\[
\revisioncolor
 \left(s_{g,i}^{+}\right)^2=\max\!\left[\sigma_g^2(d_i)-
 \frac{\operatorname{Cov}\{G(d_i),\bar Y_g(q)\mid D_t\}^2}
 {\operatorname{Var}\{\bar Y_g(q)\mid D_t\}},10^{-12}\right].
\]
Thus every hypothetical standardized toxicity margin is an affine line in the same scalar variable,
\begin{equation}\label{eq:ckg_gate_lines}
 z_{g,i}^{+}(Z_g)=a_i+c_iZ_g,\qquad
 a_i=\frac{g^\dagger-\mu_{g,i}}{\rev{s_{g,i}^{+}}},\quad
 c_i=-\frac{b_{g,i}}{\rev{s_{g,i}^{+}}}.
\end{equation}
Write $q_\tau=\Phi^{-1}(\tau)$, and let $\varphi$ denote the standard normal density. The finite evaluator follows four steps.
\begin{enumerate}
\item The crossings of $a_i+c_iZ_g=q_\tau$ partition the real line into intervals on which the set $S=\{i:z^+_{g,i}(Z_g)>q_\tau\}$ is fixed.
\item On an interval with $S\ne\varnothing$, the conditional value is
$\mathbb E_{Z_e}[\max_{i\in S}\{\mu_{e,i}+b_{e,i}Z_e\}]$. Sorting slopes and removing dominated lines gives its upper-envelope segments. If line $i$ is active on $(\ell,u)$, its contribution is
\begin{equation}\label{eq:ckg_envelope_integral}
 \mu_{e,i}\{\Phi(u)-\Phi(\ell)\}+b_{e,i}\{\varphi(\ell)-\varphi(u)\}.
\end{equation}
\item On an interval with $S=\varnothing$, pairwise crossings of the lines in Equation~\eqref{eq:ckg_gate_lines} partition the greatest-$z_g$ fallback. If candidate $j$ is selected on a segment, independence gives $\mathbb E_{Z_e}(\mu_{e,j}+b_{e,j}Z_e)=\mu_{e,j}$.
\item Each conditional value is multiplied by its standard-normal $Z_g$ interval probability and the pieces are summed. Subtracting $V_t$ gives $\alpha_{\mathrm{cKG}}(q)$; centering the efficacy intercepts at $V_t$ before integration avoids subtracting large offset-shifted values.
\end{enumerate}
The final-selection value is evaluated over all currently available combinations in $\mathcal G_t$; $\mathcal E_t$ restricts the next assignment but not the combinations entering that value. Conditional on the $10^{-12}$ variance floor, finite $\mathcal G_t$, independent Gaussian response channels, and fixed hyperparameters during the hypothetical update, the interval decomposition computes the one-cohort score deterministically, subject only to floating-point arithmetic. It does not project later cohorts, future dose availability, or stopping.

\subsection{Numerical validation}\label{sec:numerical_impl}

We checked the deterministic cKG calculation on 720 posterior states comprising 18,000 query evaluations. On a stratified 64-state subset, eight independently scrambled Sobol approximations per state (512 checks in total), each using 131,072 points, differed from the deterministic scores by at most $9.40\times10^{-5}$.

For the entropy rule, we used nine-point Gauss--Hermite quadrature, which approximates a Gaussian expectation by a weighted sum at selected integration points. A finer 128-point calculation gave the same dose choices in the checked cases where at least one combination met the toxicity criterion. This comparison does not establish a general integration error bound.

\rev{We independently checked the sensitivity archive's observations and analyses.} All 24,000 paths and their 72,000 nested snapshots were recovered. Recorded random seeds reproduced the Gaussian outcome innovations and initialization locations. Dense Mat\'ern-$5/2$ conditioning with each snapshot's saved fitted parameters reproduced its posterior predictions; the largest absolute discrepancy in a posterior mean was below $4\times10^{-9}$. Recomputed means, MCSEs, conditional denominators, and interval endpoints agreed with all 118 primary and 1,227 secondary summaries. These are checks of recorded paths and fixed-parameter calculations, not new adaptive trials or independent GP hyperparameter refits.

\rev{We compared maintained software candidate 0.2.1rc4 with the frozen study implementation in one numerical environment, using 31 prospectively specified matched configurations.} Shared posterior, score, allocation, and outcome fields agreed within the prespecified numerical tolerances in 30 pairs, including all 24 full-budget configurations drawn from the paper settings. \rev{The candidate corrects cancellation when computing normal probabilities over narrow intervals.} Relative to the preceding candidate (0.2.1rc3), all 31 configurations retained identical fitted parameters, allocation paths, and terminal results; the largest absolute cKG score change was $4.44\times10^{-16}$. In one short-budget gradual-availability tMSE pair, final selection changed because the maintained implementation restricts that choice to combinations already opened; the fitted models and allocation path were identical. This boundary correction does not change the reported full-budget design, whose grid is fully opened before completion; stopped trials make no final selection. \rev{This comparison checks software maintenance; it does not re-estimate operating characteristics.}

\rev{The maintained candidate also validates inputs and controller state, reports final-selection status explicitly, and versions checkpoint keys. The reported 40-record gradual-availability simulations open the entire grid before scheduled completion, so the restriction above does not affect their results. Reported operating characteristics refer to the archived study source. Clinical use requires a separately specified and validated complete trial design.}

\Needspace{8\baselineskip}
\subsection{\rev{Fixed-data computation times}}\label{sec:runtime}

\rev{We benchmarked the installed pre-rename software candidate 0.2.1rc4 on three prespecified, nested OSA stratum-0 datasets with 4, 20, and 40 on-grid observations, generated nonadaptively with fixed seed 20261001. Each dataset used the same two fitted GPs for all four rules, with 25 available combinations, $\tau=0.7$, and cohort size two. Table~\ref{tab:fixed_data_runtime} reports the median of five scoring repetitions after one warm-up. Scoring includes construction of the gate and shared context, with a fresh context and no score reuse; fitted-GP prediction caches were warm. GP fitting was measured once per dataset after a separate warm-up pair, with 120 Adam steps per response. Execution was serial on an Apple M4 CPU with 16 GB memory and numerical-library and PyTorch thread limits set to one. The benchmark script, fixed observations, scores and individual timings are retained alongside the maintained software source. These timings measure fixed-data operations and exclude adaptive trial simulation. The current package is named \texttt{dose-combination-bo}; renaming it does not change the numerical implementation.}

\begin{table}[H]
\revisioncolor
\centering\footnotesize
\caption{Computation times for the installed software on fixed OSA datasets.}
\label{tab:fixed_data_runtime}
\begin{tabular}{@{}rrrrrrr@{}}
\toprule
& & & \multicolumn{4}{c}{Median scoring time (ms)} \\
\cmidrule(lr){4-7}
Records & \shortstack{Gate-admissible\\combinations} & \shortstack{Two-GP\\fitting (s)} & cKG & tMSE & Entropy & cEI\\
\midrule
4 & 25 & 0.136 & 112.0 & 0.423 & 12.90 & 14.95 \\
20 & 10 & 0.136 & 104.7 & 0.446 & 13.03 & 14.87 \\
40 & 12 & 0.147 & 113.9 & 0.421 & 13.21 & 15.08 \\
\bottomrule
\end{tabular}
\vspace{3pt}
\begin{minipage}{0.98\linewidth}\scriptsize\raggedright
\textit{Note.} Each rule returns a score vector on the 25-combination grid. Fitting times are single measurements for the pair of response models, rather than medians or full-trial runtimes.
\end{minipage}
\end{table}

\subsection{Simulation settings and source context}

\paragraph{Obstructive sleep apnea (OSA)-derived simulation specification.} The OSA-derived setting uses
standardized dose $d=(d_1,d_2)\in[0,1]^2$, stratum $z\in\{0,1\}$, and the \rev{seven-term column basis
$b(d)=(1,d_1,d_2,d_1d_2,d_1^2,d_2^2,d_1^2d_2^2)^\top$}. The implementation follows the polynomial specification in archived version~1 of \citet{willardobd}, with $f=-e$, latent toxicity score $g$, and the rounded coefficient vectors reported there:
\[
 f(d,z)=\beta_z^\top b(d),\qquad g(d,z)=\theta^\top b(d),
\]
\[
\begin{aligned}
\beta_0 &=(-1.38,-4.08,-0.48,-4.23,2.45,-7.51,-1.56)^{\rev{\top}},\\
\beta_1 &=(1.05,-11.28,-8.32,-17.02,8.17,2.34,4.61)^{\rev{\top}},\\
\theta &=(-0.59,1.83,2.26,-4.05,1.79,0.47,2.91)^{\rev{\top}}\quad\text{(shared)}.
\end{aligned}
\]
The toxicity thresholds are $g^\dagger(0)=1.5$ and $g^\dagger(1)=2.0$; fixed within-combination SDs are $\sigma_{ye}=7.68$ and $\sigma_{yg}=1.29$ in both strata. On a fixed $200\times200$ grid, the continuous constrained optima under these rounded coefficients are approximated by $(0.337,0.819)$ and $(0.543,0.874)$ and lie numerically on the toxicity boundary. The thresholds and outcome SDs are simulator inputs, not clinically calibrated cutoffs or empirical MARIPOSA estimates.

\paragraph{Source context.} \citet{willardobd} constructed the OSA-derived simulator from published arm-level changes in the 4\%-desaturation apnea--hypopnea index (AHI4) and adverse-event frequencies, using unvalidated weights to form a synthetic burden score. \rev{The strata represent mild/moderate ($z=0$) and severe ($z=1$) OSA. Toxicity inherits the natural-log scale $\log(B+0.5)$, where $B$ is the weighted adverse-event burden. The present simulator generates Gaussian responses with polynomial mean $g(d,z)$ on this scale; it does not generate individual adverse-event burdens.} No patient-level records were fitted here. In MARIPOSA, aroxybutynin varied only with atomoxetine fixed at 75 mg, so the simulator's remaining two-dimensional surface is model-generated extrapolation rather than a reconstruction of the trial \citep{schweitzer2023}.

\paragraph{Logistic and MARIPOSA-motivated simulation specifications.} The disease-agnostic logistic scenario uses \rev{$L(x)=\{1+\exp(-x)\}^{-1}$},
\[
 e(d,z)=L(-1+2.6d_1+2.2d_2-1.1d_1^2-0.9d_2^2+0.3d_1d_2),
\]
and $g(d,z)=L\{a_{0z}+a_{1z}d_1+a_{2z}d_2+a_{12,z}d_1d_2\}$, with coefficients $(-2,3,2.5,1)$ for $z=0$ and $(-2,2,1.8,0.5)$ for $z=1$, and $g^\dagger=0.30$. Both outcomes increase monotonically with either dose, efficacy is bounded on the logistic scale, and the narrow feasible region places the constrained optimum near the toxicity threshold. On an endpoint-inclusive $80\times80$ grid, the continuous-domain constrained-optimum approximations are $(0.190,0.215)$ and $(0.304,0.278)$; the within-combination efficacy SD is $0.24825$, and the toxicity-outcome SDs are $0.32188$ and $0.27135$. These probability-shaped surfaces are observed with Gaussian error; they are neither binary endpoints nor a named disease model.

The synthetic MARIPOSA-motivated surface, identical in both strata, is
\[
 e(d)=50[1-\exp\{-(1.5d_1+0.6d_2+1.8d_1d_2)\}],\qquad
 g(d)=0.5+1.4d_1+0.5d_2+0.6d_1d_2,\qquad g^\dagger=2.0.
\]
For this surface, the continuous constrained optimum is $(0.5,1.0)$, which is also on the $5\times5$ assignment grid. To see this, maximize the increasing efficacy index $1.5d_1+0.6d_2+1.8d_1d_2$ subject to $1.4d_1+0.5d_2+0.6d_1d_2\le1.5$. For $d_1\le0.5$ the best feasible $d_2$ is 1; for $d_1\ge0.5$ it is $(1.5-1.4d_1)/(0.5+0.6d_1)$, along which the index decreases. The within-combination efficacy and toxicity-outcome SDs are $19.6720$ and $0.76931$. They were chosen to give signal-to-noise ratios similar to OSA, not estimated from clinical data. The author-specified coefficients create saturating efficacy and increasing burden; they do not estimate a two-dimensional clinical surface or establish pharmacologic synergy.

\subsection{Finite-grid targets}

Table~\ref{tab:grid_targets} evaluates Equation~\eqref{eq:obd_grid} directly on the assignment grid. These deterministic values use the rounded functions and thresholds above and in Appendix~\ref{sec:demo}; they are not estimates from simulated trials. \rev{A fine evaluation grid may omit assignment doses and therefore does not define the assignment-grid target.}

\begin{table}[htbp]
\centering\small
\caption{True constrained targets on the $5\times5$ assignment grid.}
\label{tab:grid_targets}
\begin{tabular}{@{}llcccc@{}}
\toprule
Scenario & Stratum & \shortstack{Acceptable\\pairs / 25} & $d^*_{\rm grid}$ & $e(d^*_{\rm grid})$ & $g(d^*_{\rm grid})$\\
\midrule
OSA-derived & 0 & 11 & $(0.25,0.75)$ & 7.6792 & 1.2817\\
OSA-derived & 1 & 16 & $(0.50,0.75)$ & 13.2055 & 1.6223\\
Logistic & 0 & 3 & $(0.25,0.00)$ & 0.3968 & 0.2227\\
Logistic & 1 & 6 & $(0.25,0.25)$ & 0.5234 & 0.2653\\
MARIPOSA-motivated & Both & 18 & $(0.50,1.00)$ & 44.7300 & 2.0000\\
Gaussian bump & 0 & 10 & $(0.25,0.50)$ & 2.6855 & 0.3027\\
Gaussian bump & 1 & 10 & $(0.50,0.50)$ & 3.9088 & 1.7000\\
\bottomrule
\end{tabular}
\end{table}

\subsection{Outcome definitions}\label{sec:outcomes}

A combination is above the toxicity limit when its true mean satisfies $g(d,z)>g^\dagger(z)$. \rev{For above-limit final selection, the all-trial percentage divides the number of above-limit selections by all initiated trials; stopped trials contribute no above-limit selection. The selection-conditional percentage and mean efficacy use only trials that make a selection. These conditional summaries are undefined if no trial selects, and no efficacy value is assigned to stopped trials.} For the full-grid design, let $d_i$ be the combination received by post-initialization participant $i$ and define
\[
N_{\rm above}^{\rm post}=\sum_{i=1}^{36}\mathbf 1\{g(d_i,z)>g^\dagger(z)\}.
\]
\rev{We report $100N_{\rm above}^{\rm post}/36$, the percentage assigned above the limit after initialization, as a main allocation outcome. For gradual availability, $N_{\rm above}$ counts above-limit assignments among all enrolled participants, including the initial cohort. We report its mean per initiated trial together with enrollment and stopping. Both counts concern assignments to combinations with above-limit true mean toxicity, not observed adverse events.}

\rev{Alongside efficacy and above-limit selection, Appendix~\ref{sec:nearoptimal} reports the probability of a truly acceptable selection within a specified efficacy margin of $d^*_{\rm grid}$ in OSA. A zero margin gives exact acceptable grid-optimal selection.}

\subsection{Additional operating characteristics}\label{sec:family_results}

The main-text Table~\ref{tab:osa_absolute_oc} reports absolute OSA-derived results. Figure~\ref{fig:family_quadrant_cellwise} below shows the surface-by-$\tau$ contrasts.

\begin{figure}[htbp]\centering
\includegraphics[width=0.94\linewidth]{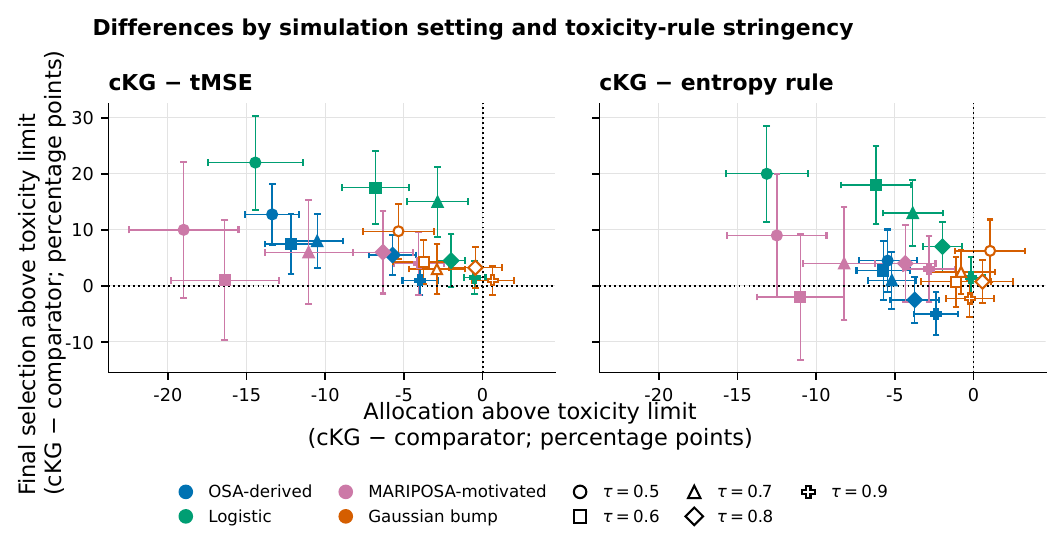}
\caption{cKG-minus-comparator differences in above-limit allocation (horizontal) and final selection (vertical) under full-grid availability. The left and right panels compare cKG with tMSE and the entropy rule, respectively. Each point is one surface--$\tau$ setting, averaged over two strata; bars are pointwise unadjusted 95\% Monte Carlo intervals. The upper-left quadrant denotes fewer above-limit assignments but more above-limit final selections under cKG. OSA-derived and Gaussian-bump settings use $M=200$ replicates; the others use $M=100$.}
\label{fig:family_quadrant_cellwise}
\end{figure}

\begin{table}[H]
\revisioncolor
\centering\footnotesize
\caption{cKG-minus-comparator differences in true mean efficacy at final selection under full-grid availability.}
\label{tab:efficacy_comparators}
\begin{tabular}{@{}llc@{}}
\toprule
Scenario & Comparator & Efficacy difference (MCSE) [95\% interval]\\
\midrule
OSA-derived & tMSE & 0.104 (0.109) [$-0.110$, 0.319] \\
OSA-derived & Entropy rule & $-0.563$ (0.119) [$-0.797$, $-0.329$] \\
Logistic & tMSE & 0.024 (0.004) [0.017, 0.031] \\
Logistic & Entropy rule & 0.024 (0.004) [0.017, 0.031] \\
MARIPOSA-motivated & tMSE & $-0.508$ (0.336) [$-1.176$, 0.159] \\
MARIPOSA-motivated & Entropy rule & 0.115 (0.463) [$-0.804$, 1.034] \\
Gaussian bump & tMSE & 0.165 (0.035) [0.096, 0.235] \\
Gaussian bump & Entropy rule & $-0.074$ (0.041) [$-0.156$, 0.007] \\
\bottomrule
\end{tabular}
\vspace{3pt}
\begin{minipage}{0.98\linewidth}\scriptsize\raggedright
\textit{Note.} Differences average two strata and five $\tau$ settings equally within each paired replicate. Intervals are pointwise, unadjusted 95\% Monte Carlo $t$ intervals. OSA-derived and Gaussian-bump scenarios use $M=200$; the others use $M=100$. Positive values favor cKG. All trials made a final selection, including fallback selections. Efficacy scales differ across scenarios.
\end{minipage}
\end{table}

\subsection{Final-selection efficacy relative to cEI}\label{sec:ckg_cei_tau}

Table~\ref{tab:ckg_cei_efficacy_tau} gives the complete $\tau$ profile for cKG versus cEI.

\begin{table}[H]
\centering
\scriptsize
\caption{cKG-minus-cEI differences in true mean efficacy at final selection, by $\tau$, under full-grid availability.}
\label{tab:ckg_cei_efficacy_tau}
\setlength{\tabcolsep}{3.5pt}
\resizebox{\textwidth}{!}{%
\begin{tabular}{@{}ccccc@{}}
\toprule
$\tau$ & OSA-derived & Logistic & \shortstack{MARIPOSA-\\motivated} & \shortstack{Gaussian\\bump} \\
\midrule
$0.5$ & $1.192\ [0.755, 1.630]$ & $0.0374\ [0.0208, 0.0539]$ & $6.160\ [5.102, 7.217]$ & $0.389\ [0.240, 0.538]$ \\
$0.6$ & $1.034\ [0.622, 1.447]$ & $0.0338\ [0.0190, 0.0486]$ & $6.188\ [5.041, 7.335]$ & $0.283\ [0.143, 0.423]$ \\
$0.7$ & $0.905\ [0.505, 1.305]$ & $0.0245\ [0.0109, 0.0381]$ & $4.172\ [2.756, 5.589]$ & $0.174\ [0.054, 0.295]$ \\
$0.8$ & $0.391\ [0.040, 0.743]$ & $0.0091\ [-0.0024, 0.0206]$ & $3.463\ [2.007, 4.919]$ & $0.044\ [-0.043, 0.130]$ \\
$0.9$ & $0.172\ [-0.125, 0.468]$ & $0.0080\ [-0.0013, 0.0173]$ & $3.092\ [1.534, 4.649]$ & $0.063\ [-0.022, 0.147]$ \\
\bottomrule
\end{tabular}%
}
\vspace{3pt}
\begin{minipage}{0.98\linewidth}
\scriptsize\raggedright
\textit{Note.} Entries are mean differences [pointwise unadjusted 95\% Monte Carlo $t$ interval], weighting the two strata equally; positive values indicate higher true mean efficacy under cKG. OSA-derived and Gaussian-bump analyses use $M=200$ replicates, and the others use $M=100$. Efficacy scales differ by setting. Results include prespecified fallback selections; in the Gaussian-bump analysis, these occurred in 17.25--41.75\% of cKG trials and 17.25--41.25\% of cEI trials across $\tau$.
\end{minipage}
\end{table}

\subsection{Whether a final selection passed the toxicity criterion}\label{sec:selection_partition}

Table~\ref{tab:selection_partition} partitions the original final selections without changing or rerunning either policy. These exploratory summaries distinguish four mutually exclusive categories: criterion-passing and truly acceptable, criterion-passing and truly above limit, fallback and truly acceptable, or fallback and truly above limit. The categories use all initiated full-grid trials as the denominator. Their sum is 100\% within each row because these runs always made a final selection. \rev{Adding a no-selection option would require specifying its value, revising the lookahead, and running new adaptive simulations. It cannot be evaluated by relabeling these recorded selections.}

\begin{table}[H]
\centering
\footnotesize
\setlength{\tabcolsep}{3pt}
\caption{Final-selection status and true mean toxicity for cKG and cEI (percent of all trials).}
\label{tab:selection_partition}
\begin{tabular}{@{}llcccc@{}}
\toprule
& & \multicolumn{2}{c}{Passed posterior criterion} & \multicolumn{2}{c}{Fallback selection} \\
\cmidrule(lr){3-4}\cmidrule(lr){5-6}
Scenario & Rule & Acceptable & Above limit & Acceptable & Above limit \\
\midrule
OSA-derived & cKG & 80.55 & 18.55 & 0.65 & 0.25 \\
& cEI & 88.30 & 10.15 & 1.20 & 0.35 \\
Logistic & cKG & 71.70 & 21.80 & 5.20 & 1.30 \\
& cEI & 81.70 & 10.70 & 5.70 & 1.90 \\
MARIPOSA-motivated & cKG & 79.60 & 20.40 & 0.00 & 0.00 \\
& cEI & 99.80 & 0.20 & 0.00 & 0.00 \\
Gaussian bump & cKG & 63.70 & 9.40 & 16.75 & 10.15 \\
& cEI & 62.95 & 10.55 & 16.80 & 9.70 \\
\bottomrule
\end{tabular}
\vspace{3pt}
\begin{minipage}{0.98\linewidth}
\scriptsize\raggedright
\textit{Note.} Each row averages two strata and five $\tau$ values equally. Acceptable means $g(\hat d,z)\le g^\dagger(z)$ in the simulation truth, not a demonstrated clinical safety claim. Counts are 2,000 trials per policy in OSA and Gaussian bump and 1,000 in each other scenario. These are descriptive post hoc partitions of the same records used for the main contrasts; they are not new simulation experiments.
\end{minipage}
\end{table}

\rev{Machine-readable files include conditional efficacy and the corresponding denominators. Because these subsets differ across policies, their summaries do not measure unconditional policy effects.} In Gaussian bump, for example, criterion-passing selections have mean efficacy 1.433 under cKG (1,462 trials) and 1.189 under cEI (1,470 trials), whereas fallback selections have means 0.486 (538 trials) and 0.430 (530 trials). Reporting only the criterion-passing means would omit more than one quarter of the original trials.

\subsection{\rev{Acceptable selection within an efficacy margin}}\label{sec:nearoptimal}

\rev{Using the recorded full-grid OSA selections, we estimated the probability of selecting a combination with true mean toxicity at or below its limit and true mean efficacy within $\delta$ of $e(d^*_{\rm grid},z)$. The reference is the best truly acceptable combination on the 25-point assignment grid, not a continuous-domain optimum. Before deriving these post hoc summaries, we fixed $\delta\in\{0,0.5,1,2\}$ simulated AHI4 events/hour as a descriptive sensitivity range, without clinical calibration. All initiated trials contribute, retaining fallback selections and evaluating them by the same truth-based criterion.}

\rev{The $0.5$- and $1$-event margins give the same rates as zero because neither stratum has another acceptable grid combination within one event/hour of its optimum. cKG had higher estimated rates than cEI at every cutoff and margin. Entropy had higher estimated rates than cKG throughout; the cKG--tMSE ranking depended on the cutoff and margin. These summaries complement mean efficacy and above-limit selection; they do not establish a clinical benefit--risk ranking.}

\begin{table}[H]
\revisioncolor
\centering
\footnotesize
\setlength{\tabcolsep}{5pt}
\caption{Truly acceptable selections within an efficacy margin of the grid optimum in OSA, under full-grid availability.}
\label{tab:nearoptimal}
\begin{tabular}{@{}clcccc@{}}
\toprule
$\tau$ & Rule & $\delta=0$ & $\delta=0.5$ & $\delta=1$ & $\delta=2$ \\
\midrule
0.5 & cKG & 21.00 (2.08) & 21.00 (2.08) & 21.00 (2.08) & 38.00 (2.41) \\
0.5 & tMSE & 24.75 (2.30) & 24.75 (2.30) & 24.75 (2.30) & 46.25 (2.58) \\
0.5 & Entropy rule & 22.25 (2.09) & 22.25 (2.09) & 22.25 (2.09) & 41.75 (2.62) \\
0.5 & cEI & 17.00 (1.99) & 17.00 (1.99) & 17.00 (1.99) & 36.00 (2.63) \\
\addlinespace[2pt]
0.6 & cKG & 18.25 (2.04) & 18.25 (2.04) & 18.25 (2.04) & 34.50 (2.41) \\
0.6 & tMSE & 19.50 (1.97) & 19.50 (1.97) & 19.50 (1.97) & 36.75 (2.42) \\
0.6 & Entropy rule & 21.00 (2.05) & 21.00 (2.05) & 21.00 (2.05) & 38.50 (2.50) \\
0.6 & cEI & 10.25 (1.52) & 10.25 (1.52) & 10.25 (1.52) & 30.25 (2.60) \\
\addlinespace[2pt]
0.7 & cKG & 14.25 (1.71) & 14.25 (1.71) & 14.25 (1.71) & 28.75 (2.23) \\
0.7 & tMSE & 13.75 (1.81) & 13.75 (1.81) & 13.75 (1.81) & 30.50 (2.50) \\
0.7 & Entropy rule & 14.50 (1.86) & 14.50 (1.86) & 14.50 (1.86) & 32.00 (2.48) \\
0.7 & cEI & 7.25 (1.39) & 7.25 (1.39) & 7.25 (1.39) & 21.00 (2.08) \\
\addlinespace[2pt]
0.8 & cKG & 9.75 (1.61) & 9.75 (1.61) & 9.75 (1.61) & 23.25 (2.06) \\
0.8 & tMSE & 11.00 (1.59) & 11.00 (1.59) & 11.00 (1.59) & 18.00 (2.13) \\
0.8 & Entropy rule & 11.00 (1.59) & 11.00 (1.59) & 11.00 (1.59) & 24.25 (2.18) \\
0.8 & cEI & 7.00 (1.28) & 7.00 (1.28) & 7.00 (1.28) & 17.25 (1.99) \\
\addlinespace[2pt]
0.9 & cKG & 5.75 (1.29) & 5.75 (1.29) & 5.75 (1.29) & 13.50 (1.83) \\
0.9 & tMSE & 5.50 (1.22) & 5.50 (1.22) & 5.50 (1.22) & 10.75 (1.66) \\
0.9 & Entropy rule & 9.75 (1.53) & 9.75 (1.53) & 9.75 (1.53) & 17.25 (1.99) \\
0.9 & cEI & 5.50 (1.22) & 5.50 (1.22) & 5.50 (1.22) & 12.00 (1.63) \\
\bottomrule
\end{tabular}
\vspace{3pt}
\begin{minipage}{0.98\linewidth}
\scriptsize\raggedright
\textit{Note.} Entries are percentages (MCSE in pp), averaging the two stratum indicators within each of $M=200$ independent replicate sets. Each rule--cutoff row uses all 400 initiated trials, retaining fallback selections. Acceptability uses true mean toxicity, not the posterior criterion. Margins are in simulated AHI4 events/hour and are descriptive, not clinically calibrated. $\delta=0$ denotes selection of the exact acceptable grid optimum. These post hoc summaries reuse existing records; no adaptive trials were rerun.
\end{minipage}
\end{table}

\subsection{Dose availability and stopping sensitivity}\label{sec:protocolscaffold}

The OSA-derived gradual-availability analysis is motivated by \citet{willardobd} but is not a reconstruction. The two strata are analyzed separately; no shared accrual, cross-stratum borrowing, or enrollment reallocation is modeled.

\rev{The gradual design begins with two participants at $(0,0)$. After cohort $t$, the grid available for the next cohort is}
\[
\revisioncolor
 \mathcal G_t=\{d\in\mathcal G:d_1+d_2\le0.25(t+1)\}.
\]
\rev{The full grid is available before the eighth cohort.} Before then, previously assigned combinations are ineligible while an available unassigned combination remains. On the first two consecutive occasions when no eligible combination meets $\PF(d)>\tau$, Equation~\eqref{eq:si_emptygate} supplies the fallback; a third such occasion stops the simulated trial before assignment and without a final selection. The count resets when an eligible combination meets the requirement. At maximum enrollment, the final rule uses the full grid. Dose availability follows this fixed schedule rather than observed outcomes. Permanent elimination, delayed outcomes, independent monitoring, and acquisition-value stopping are not modeled.

The analysis includes both strata, the four reported rules, $M=200$ independent replicates, and $\tau=0.7,0.9$. Within a replicate, rules share random-number streams for initialization and outcome errors; each rule--$\tau$ setting therefore contains 400 initiated stratum-specific trials. Selection-conditional contrasts include only matched pairs in which both rules make a final selection; other contrasts include all matched pairs. Intervals are pointwise 95\% Monte Carlo $t$ intervals. Main-paper Table~\ref{tab:gradual} reports the operating characteristics and denominators.

\section{Sensitivity to participant count and assumed toxicity-outcome variability}\label{sec:noiserobust}

\subsection{Scope, paths, and snapshot estimands}

The nested sensitivity analysis uses the OSA-derived surface in two strata, the full $5\times5$ dose set, $\tau\in\{0.7,0.9\}$, four \rev{initialization observations}, cohorts of two, and no stopping. It compares cKG with tMSE, the entropy rule, and cEI.

Efficacy and toxicity outcomes are generated with fixed within-combination SDs of 7.68 and 1.29, respectively. The analysis model uses 7.68 for efficacy and 0.645, 1.29, or 2.58 for toxicity. \rev{These SDs remain fixed throughout each simulated trial.} Thus only the assumed toxicity-outcome SD changes, through
\[
c_\sigma=\frac{\text{toxicity-outcome SD specified in the analysis}}
{\text{toxicity-outcome SD used to generate the data}}\in\{0.5,1,2\}.
\]
The three values understate, match, or overstate within-combination toxicity-outcome variability. The data-generating surfaces and outcome SDs, efficacy model, and all other analysis choices remain fixed. Settings share Gaussian innovations within replicate, although adaptive assignments can produce different dose--outcome histories. This analysis therefore isolates one likelihood assumption and does not address binary DLTs or other forms of model misspecification.

Each simulated trial continues to 80 response records and is analyzed after $R\in\{20,40,80\}$ records. Because four records are \rev{initialization observations}, the analyses follow $R-4\in\{16,36,76\}$ participant assignments. \rev{At snapshot $R$, let $N_{\rm above}^{\rm post}$ count above-limit assignments among those $R-4$ participants, and let $\hat d_R$ be the combination selected using the first $R$ records. Allocation is summarized as $100N_{\rm above}^{\rm post}/(R-4)$. Above-limit final selection is reported as the percentage of trials with $g(\hat d_R)>g^\dagger$.}
\rev{The response models use the initialization records, but allocation summaries exclude them.} These participant counts are not clinical sample-size recommendations; $M$ denotes independent simulation replicates, not enrollment.

\subsection{Contrast families and uncertainty rules}

Each outcome is averaged equally over the two strata and two $\tau$ settings within replicate. The displayed family contains 54 cKG-minus-comparator contrasts: three comparators, two outcomes, three snapshots, and three $c_\sigma$ settings. The replicate is the independent Monte Carlo unit. Simultaneous 95\% max-$t$ intervals use the higher empirical 0.95 quantile of the maximum absolute studentized coordinate from 100,000 Rademacher multiplier draws.

Replication began at $M=200$ and all settings were extended together to $M=500$ under a prespecified precision screen \citep{morris2019,koehler2009,burton2006}. The cap of 500 was set before the $M=500$ effects were examined. At that cap, 51 of the 54 displayed contrasts met the \rev{1.5-pp} MCSE criterion. \rev{The three exceptions were above-limit final-selection contrasts with overstated toxicity variability: cKG versus the entropy rule after 20 and 40 records and cKG versus cEI after 20 records; the largest MCSE was 1.744 pp.} The complete 118-row precision screen is retained in the reproducibility files. The analysis does not establish a clinical sample size or robustness to other forms of model misspecification.

\subsection{Acquisition comparisons across snapshots and toxicity-outcome variability}

\begin{figure}[H]
\centering
\includegraphics[width=\linewidth]{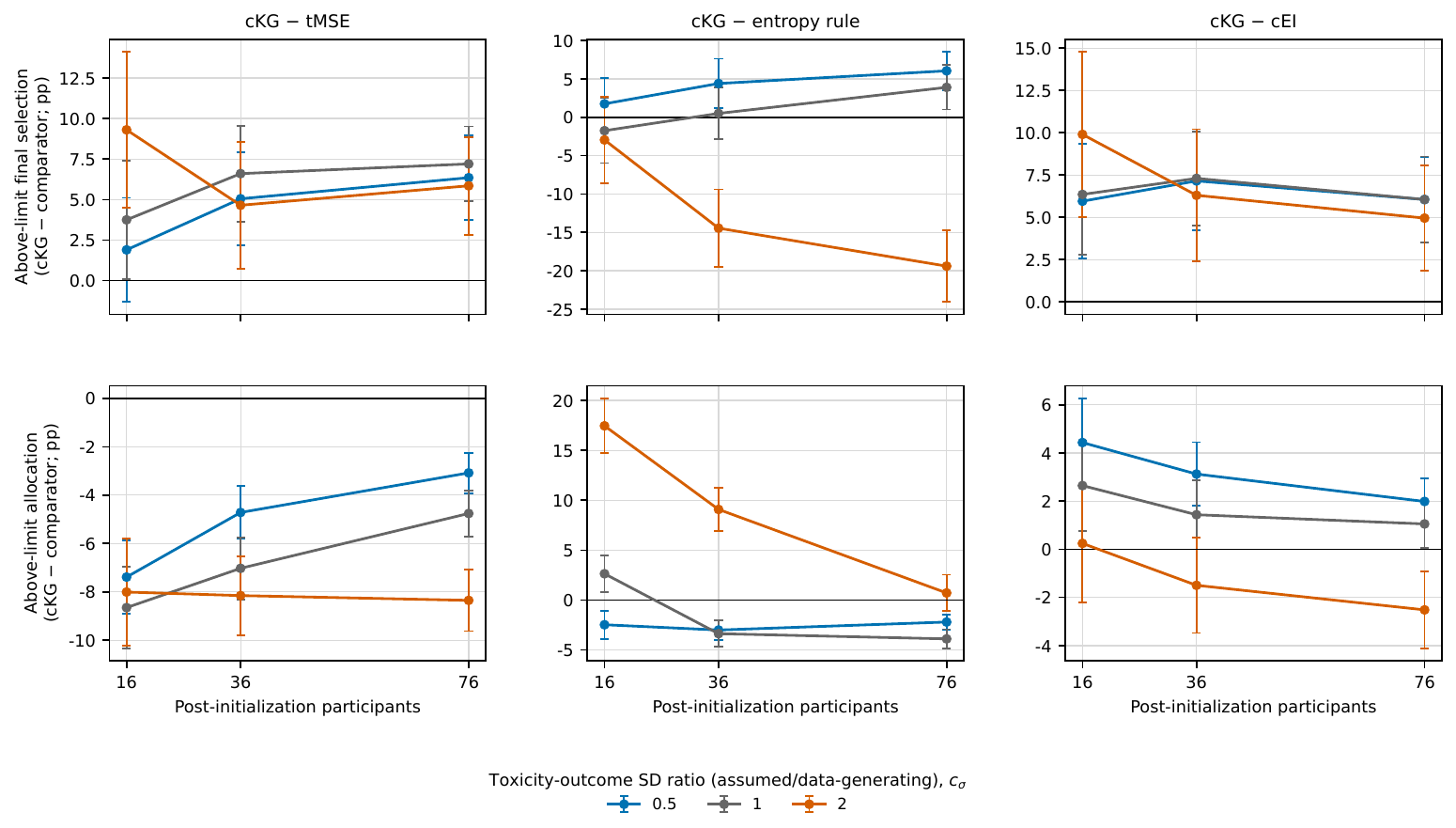}
\caption{cKG-minus-comparator differences across participant counts and assumed toxicity-outcome variability in the OSA-derived analysis ($M=500$). Columns identify the comparator; rows show above-limit final selection and post-initialization allocation. Colors show the ratio of the assumed to the data-generating toxicity-outcome SD. Points weight the two strata and $\tau=0.7,0.9$ equally; this average is descriptive. The 16, 36, and 76 assignments correspond to 20, 40, and 80 response records. Bars are simultaneous 95\% max-$t$ Monte Carlo intervals across the 54 contrasts. Positive values indicate more above-limit final selections or assignments under cKG.}
\label{fig:calibration-profile}
\end{figure}

Figure~\ref{fig:calibration-profile} displays all 54 cKG-minus-comparator contrasts summarized in the main paper.

\rev{Table~\ref{tab:nested_efficacy} reports final-selection efficacy for the same nine SD-by-snapshot settings. These are descriptive means over all trials, including fallback selections; there was no stopping. These summaries are separate from the family of 54 toxicity contrasts.}

\begin{table}[H]
\revisioncolor
\centering\footnotesize
\caption{Mean true efficacy at final selection across assumed toxicity SDs and record counts.}
\label{tab:nested_efficacy}
\begin{tabular}{@{}cccccc@{}}
\toprule
$c_\sigma$ & Records $R$ & cKG & tMSE & Entropy & cEI\\
\midrule
0.5 & 20 & 6.43 (0.12) & 6.60 (0.12) & 6.89 (0.11) & 6.12 (0.11) \\
0.5 & 40 & 6.52 (0.11) & 6.45 (0.10) & 6.68 (0.10) & 5.94 (0.10) \\
0.5 & 80 & 6.68 (0.10) & 6.46 (0.09) & 6.75 (0.09) & 5.98 (0.09) \\
1 & 20 & 6.56 (0.12) & 6.34 (0.11) & 7.19 (0.12) & 5.98 (0.11) \\
1 & 40 & 6.73 (0.10) & 6.49 (0.10) & 7.24 (0.11) & 6.15 (0.10) \\
1 & 80 & 7.08 (0.10) & 6.79 (0.09) & 7.34 (0.10) & 6.24 (0.10) \\
2 & 20 & 6.98 (0.15) & 6.01 (0.14) & 8.68 (0.15) & 6.38 (0.14) \\
2 & 40 & 6.81 (0.12) & 6.44 (0.12) & 8.90 (0.15) & 6.49 (0.12) \\
2 & 80 & 6.92 (0.10) & 6.92 (0.09) & 9.09 (0.13) & 6.70 (0.10) \\
\bottomrule
\end{tabular}
\vspace{3pt}
\begin{minipage}{0.98\linewidth}\scriptsize\raggedright
\textit{Note.} Entries are means (MCSEs), averaging two strata and $\tau=0.7,0.9$ equally within each of $M=500$ replicates. Four initialization records are included in $R$. The ratio $c_\sigma$ compares the assumed toxicity-outcome SD with the SD used to generate the data. \rev{Efficacy is simulated AHI4 reduction in events/hour.} These marginal summaries do not provide paired uncertainty for differences between rules.
\end{minipage}
\end{table}

\section{Gaussian-bump simulation specification}\label{sec:demo}

\begin{figure}[t]\centering
\includegraphics[width=\linewidth]{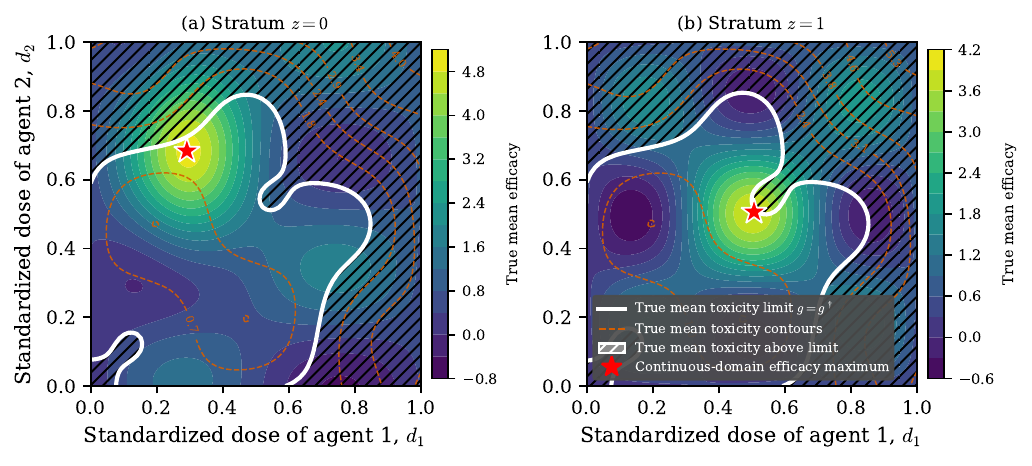}
\caption{Gaussian-bump true mean efficacy and toxicity by stratum. Filled contours show efficacy, dashed orange contours show toxicity, the white contour marks $g=g^\dagger$, and hatching marks combinations above the limit. The red star is the fine-grid approximation to the continuous-domain unconstrained efficacy maximum. Doses are standardized to $[0,1]$; color scales are panel-specific, and the $5\times5$ candidate grid is omitted.}
\label{fig:simsurf}
\end{figure}

\rev{The Gaussian-bump scenario has nonmonotone mean toxicity and Gaussian-shaped peaks in efficacy.} Its continuous-domain unconstrained efficacy maximum has mean toxicity below the limit in both strata; on the $5\times5$ assignment grid this holds only in stratum~1. \rev{Each stratum has 10 acceptable combinations on the assignment grid. In stratum~0, mean toxicity increases from 0.303 at $(0.25,0.50)$ to 1.250 at the adjacent pair $(0.50,0.50)$. The lowest pair $(0,0)$ exceeds the limit in both strata; these full-grid runs do not use it as a required initial assignment.} Figure~\ref{fig:simsurf} shows the geometry. The same full-grid algorithms and outcome measures are used, with $M=200$ replicates and one simulated trial per stratum in each replicate.

The exact data-generating functions are
\[
\revisioncolor
\begin{aligned}
e(d,0)&=3.2\exp\{-[(d_1-0.3)^2/0.03+(d_2-0.7)^2/0.1]\}\\
&\quad+\sin(2\pi d_1)\cos(3\pi d_2)+1.4d_1\exp(-1.2d_1^2)+0.9d_2\exp(-0.6d_2^2),\\
e(d,1)&=2\exp\{-[(d_1-0.5)^2+(d_2-0.5)^2]/0.05\}\\
&\quad+1.2\sin(3\pi d_1)\sin(3\pi d_2)+d_1\exp(-d_1^2)+0.8d_2\exp(-0.9d_2^2),\\
g(d,0)&=3[(d_1-0.5)^2+(d_2-0.5)^2]+0.5\sin(3\pi d_1)\sin(3\pi d_2)+1.5(d_1^2+d_2^2),\\
g(d,1)&=4[(d_1-0.5)^2+(d_2-0.5)^2]+0.7\sin(3\pi d_1)\sin(3\pi d_2)+2(d_1^2+d_2^2).
\end{aligned}
\]
On the endpoint-inclusive $80\times80$ evaluation grid, the median toxicity defines $g^\dagger=1.29633$ and $1.73722$ in strata 0 and 1. The corresponding continuous-domain constrained-optimum approximations are $(0.291,0.684)$ and $(0.506,0.506)$. Efficacy-outcome SDs are $1.86599$ and $1.48427$; toxicity-outcome SDs are $1.19728$ and $1.60591$. These fixed values are not empirical disease-specific variability estimates. \rev{Because several surface features differ from OSA, these comparisons cannot isolate the effect of multimodality or any other single feature.}

\end{document}